\documentclass[pre,twocolumn,superscriptaddress,floatfix]{revtex4-1}

\usepackage{amsmath}
\usepackage{graphicx}

\usepackage[multidot]{grffile}

\usepackage{booktabs}
\usepackage{microtype}
\usepackage[colorlinks,bookmarks=false,citecolor=blue,linkcolor=red,urlcolor=blue]{hyperref}

\usepackage{mdframed}
\mdfdefinestyle{algstyle}{innertopmargin=5, innerbottommargin=5,
  innerleftmargin=0,  innerrightmargin=0,
  leftmargin=-10,
hidealllines=true}

\usepackage{algpseudocode}
\usepackage{dsfont}

\begin{document}
\title{Valence-Bond Quantum Monte Carlo Algorithms Defined on Trees}
\author{Andreas Deschner}
\email{deschna@mcmaster.ca}
\author{Erik S. S{\o}rensen}
\email{sorensen@mcmaster.ca} \affiliation{Department of Physics and Astronomy, McMaster University, Hamilton,
Canada L8S 4M1}

\begin{abstract}
  We present a new class of algorithms for performing valence-bond
  quantum Monte Carlo of quantum spin models. Valence-bond quantum Monte
  Carlo is a projective $T$\hspace{0pt}$=$\hspace{0pt}$0$ Monte Carlo
  method based on sampling of a set of operator-strings that can be viewed
  as forming a tree-like structure.  The algorithms presented here
  utilize the notion of a worm that moves up and down this tree and
  changes the associated operator-string.  In quite general terms we
  derive a set of equations whose solutions correspond to a new class of
  algorithms.  As specific examples of this class of algorithms we focus
  on two cases.  The \emph{bouncing worm} algorithm, for which updates
  are \emph{always accepted} by allowing the worm to bounce up and down
  the tree and  the \emph{driven worm} algorithm, where a single
  parameter controls how far up the tree the worm reaches before turning
  around. The latter algorithm involves only a \emph{single bounce}
  where the worm turns from going up the tree to going down.  The
  presence of the control parameter necessitates the introduction of an
  acceptance probability for the update.   
\end{abstract} 

\maketitle

\section{Introduction}
\label{sec_intro}
Projective techniques are often used for determining the ground-state
properties of strongly correlated models defined on a lattice. They were
initially developed for non-lattice models~\cite{Ceperley_1986} and then
used for the study of fermionic lattice
models~\cite{Blankenbecler_1983}. They were subsequently applied to
quantum spin
models~\cite{liang_new_1988,*liang_existence_1990,liang_monte_1990,
Trivedi_1990,Runge_1992,calandra_charge_1998,capriotti_long_range_1999}
as well as other models.  The underlying idea is easy to describe. For a
lattice Hamiltonian $H$, it is possible to choose a constant $c$ such
that the dominant eigenvalue $E$ of $c\mathds{1}-H$ corresponds to the
ground-state wavefunction of $H$, $|\Psi_0\rangle$. We can then use
$P=c\mathds{1}-H$ as a projective operator in the sense that the
repeated application of $P$  to a trial wave function,
$P^n|\Psi_T\rangle$, will approach $E^n|\Psi_0\rangle$ for large $n$.
Hence, if $n$ can be taken large enough, $|\Psi_0\rangle$ can be
projected out in this manner provided that
$\langle\Psi_0|\Psi_T\rangle\neq 0$.  Some variants of this approach are
often referred to as Green's functions Monte Carlo
(GFMC)~\cite{Blankenbecler_1983,Trivedi_1990,Runge_1992,calandra_charge_1998,capriotti_long_range_1999}.
Other projective operators such as $\exp(-\tau H)$ can be used depending
on the model and its spectrum. For a review see
Ref.~\onlinecite{Linden_1992,Cerf_1995,Sorella_2007}.  The convergence
of such projective techniques may be non-trivial as can be shown by
analyzing simple models~\cite{Hetherington_1984}.  If $P|\Psi_T\rangle$
can be evaluated exactly, this projective scheme 
is equivalent to the power method as used in exact diagonalization studies.
As the number of
sites in the lattice model is increased, exact evaluation quickly
becomes impossible and Monte Carlo methods (projector Monte Carlo) have
to be used.

The efficiency of the Monte Carlo sampling is crucial for the
performance of implementations of the projective method and detailed
knowledge of such Monte Carlo methods is of considerable importance.
Here, we have investigated a new class of Monte Carlo algorithms for
projective methods for lattice models.  We discuss these algorithms
within the context of quantum Monte Carlo where the projection is
performed in the valence bond
basis~\cite{liang_new_1988,*liang_existence_1990,liang_monte_1990,
sandvik_ground_2005, *beach_formal_2006, *sandvik_monte_2007,
sandvik_loop_2010}, so called valence bond quantum Monte Carlo (VBQMC).
The algorithms are, however, applicable to projective techniques in any
basis.

VBQMC was first developed by Liang~\cite{liang_new_1988,
*liang_existence_1990, liang_monte_1990} and then, starting fifteen
years later, significantly further developed by Sandvik and
collaborators~\cite{sandvik_ground_2005,
*beach_formal_2006,*sandvik_monte_2007, sandvik_loop_2010} and it is now
widely used.  Since its inception, VBQMC has been improved and
generalized in several ways: it can be used on systems with spins with
$S\neq1/ 2$~\cite{liang_monte_1990} and states with total $S_z = 1/
2$~\cite{banerjee_generalization_2010}.  An efficient sampling algorithm
with loop updates is known for systems with
$S=1/2$~\cite{sandvik_loop_2010}.  

\begin{figure}[t]
  \includegraphics[width=.85\linewidth]{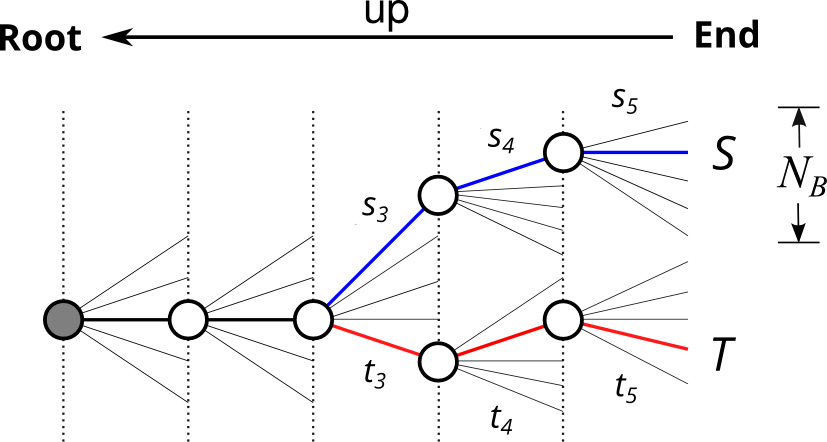}
  \caption{The branching tree of length 5 for the selection of an
    operator string in a system with a Hamiltonian of $N_B$ (here 5)
    terms. The operator that acts on the state first, is chosen at the
    first node on the left. This node is called the root and the
    direction towards the root we define to be up. The operator that
    acts on the state last is at the end of the string.  The two colored
    paths differ in the choice of the last three operators.  The last
    three branches, thus, contribute different operators and weights
    ($s_i$, $t_i$). The resulting strings \textit{\textsf{S}} and
    \textit{\textsf{T}} are different.
  }
  \label{fig_tree}
\end{figure}
As outlined above, VBQMC works by projecting onto the ground-state by repeatedly
acting on a trial-state $|\Psi_T\rangle$ with $P = c\mathds{1}-H$, where
the constant $c$ is chosen such that the ground-state has the biggest
eigenvalue. For Hamiltonians with bounded spectrum such a $c$ can always
be found. For a simple quantum spin model defined on a lattice we have 
\begin{equation}
H=J\sum_{<i,j>}\mathbf{S}_i\cdot\mathbf{S}_j=\sum_{<i,j>}h_{ij}
\end{equation}
and we can write $P = c\mathds{1}-H  = \sum O_{ij}$ as a sum over $N_B$
bond-operators $O_{ij}$.  Taking $P$ to the $n$th power then results in
a sum over products of these bond-operators $O_{ij}$:
\begin{eqnarray}
  P^n = \sum_a \underbrace{O_{i(a,1)j(a,1)} \dots
  O_{i(a,n)j(a,n)}}_{\textrm{$n$-operators}} := \sum_a S_a\;. 
  \label{eq_defopstr}
\end{eqnarray}
Each instance of this product then forms a string $S_a$ of
bond-operators of length $n$. When selecting such a string of length
$n$, one has to make a choice between the $N_B$ bond-operators at each
position in the string. It is possible to view the construction of such
a string as a specific path in a decision-tree (see
Fig.~\ref{fig_tree}).

Although the algorithms we present can be extended to higher spin
models, we shall restrict the discussion to quantum spin models with
$S=1/2$ where one usually takes $c=J N_B/4$.  The action of
the bond-operators then takes an attractively simple form.  

In a valence bond basis state spins are paired into singlets.  A
specific pairing of all spins is usually referred to as a covering.  All
such coverings form an over complete basis for the singlet sub-space of
the model. We shall only be concerned with models defined on a
bi-partite lattice in which case a given valence bond covering, $C$, for
a lattice with $N$ spins can be denoted by listing all $N/2$ pairs of
$[i,j]$ with $i$ on sub-lattice $A$ and $j$ on sub-lattice $B$.  Here,
$[i,j]=(|\uparrow_{i_A}\downarrow_{j_B}\rangle-|\downarrow_{i_A}\uparrow_{j_B}\rangle)/\sqrt{2}$.
We label the initial covering (trial-state) as $C_0$.  The action of an
operator $O_{ij}$ can take two
forms~\cite{liang_new_1988,*liang_existence_1990,liang_monte_1990,
sandvik_ground_2005}:
\begin{itemize}
  \item 
    The sites $i$ and $j$ are in a singlet before the action of the
    operator. Then, the action of the operator does not change the state and
    we can associate a weight of $w=1$:
    \begin{equation}
    O_{ij}[i,j]=1[i,j] \,.
    \end{equation}
  \item
    The sites $i$ and $j$ are not in a singlet before the action of the
    operator.
    Then, after the action of the operator, the sites $i$ and $j$ form a
    singlet.  The sites they were originally connected to are also
    returned in a singlet-state.  Furthermore, the state is  multiplied by
    a weight equal to $w=\frac{1}{2}$:
    \begin{equation}
      O_{ij}[i,k][l,j]=\frac{1}{2}[i,j][l,k] \,.
    \end{equation}
\end{itemize}
A particularly nice feature is that the application of any of the
$O_{ij}$ to any given covering yields a unique other covering and
\emph{not} a linear combination of coverings.  Although convenient, this
feature of projections in the valence bond basis is not strictly
necessary for the algorithms we discuss here as they can be adapted to
the case where a linear combination of states are
generated~\cite{SzProj_2014}.
For a given operator string $S_a=\prod_k O_k$, we can associate a weight
given by $W_a=\prod w_k$. The state $S_a C_0$ will contribute to the
final projected estimate of the ground-state with this weight. One can
then sample the ground-state by performing a random walk in the space of
all possible strings
$S_a$~\cite{liang_new_1988,*liang_existence_1990,liang_monte_1990,
sandvik_ground_2005} according to the weight $W_a$. This way of sampling
is quite different from GFMC even though VBQMC and GFMC are closely
related projective techniques. GFMC, as it is used in for instance
Ref.~\onlinecite{Runge_1992}, is usually performed in the $S^z$ basis
but can also be done in terms of the valence bond
basis~\cite{santoro_spin_liquid_1999}. In GFMC the projection is done by
stochastically evaluating the action of the whole projection-operator on
a trial-state. This is done by introducing probabilistic ``walkers''.
In contrast, as mentioned, in VBQMC a single state results and the
strings $S_a$ are sampled according to their weight.  Clearly, the
efficient sampling of states resulting from the stochastic projection of
the trial-state is a difficult problem. Here, we propose to use worm
(cluster) algorithms for this purpose.

In Monte Carlo calculations one averages over many configurations of the
system which are generated with appropriate probabilities.  Usually,
this is done in a Markov-chain, where one configuration is chosen as a
variation of the last.  One important feature of an efficient algorithm
is that these consecutive configurations are as uncorrelated as
possible. This led to the introduction of algorithms where whole
clusters and not just single elements are changed going from one
configuration to the next~\cite{swendsen_nonuniversal_1987,
evertz_cluster_1993} or where all elements in the path of a \emph{worm}
are changed~\cite{prokofev_worm_2001}.

Here, we show how it is possible to adapt such worm algorithms for
projections in the context of VBQMC\@.  The algorithms we have studied are
based on the notion of a worm moving around in the decision tree
described above. As in earlier worm algorithms, the change of many
elements is achieved by moving the worm based on local
conditions~\cite{prokofev_worm_2001, prokofev_worm_2009,
  alet_cluster_2003, alet_directed_2003, *hitchcock_dual_2004,
wang_worm_2005} and one might refer to the algorithms as tree-worm
algorithms.  In general, the algorithms can be viewed as
\emph{directed}~\cite{alet_directed_2003, *hitchcock_dual_2004}
algorithms.

When we update the string, we start with a worm at the end of the tree
and move it up the tree.  See Fig.~\ref{fig_tree}. The worm then moves
around in the tree and where it goes the operator-string is changed.
When the worm finds its way back to the bottom of the tree the update is
complete.  We derive a set of simple equations governing the movement of
the worm. The solution of these equations lead to parameters defining a
new class of algorithms. Quite generally, many solutions are possible
leaving significant room for choosing parameters that will lead to the
most optimal algorithm.

We focus on two specific choices of parameters corresponding to two
different algorithms.  The \emph{bouncing worm}  algorithm, for which
every update is accepted and the \emph{driven worm} algorithm, for which
the update is accepted with some probability. 
With the driven worm algorithm one can choose at will how much of the
operator-string is on average changed in a successful update. 

In order to test the algorithms, we calculate the ground-state energy of
the isotropic Heisenberg-chain. This quantity is easy to calculate with
VBQMC and can be exactly computed using the Bethe-ansatz. It is thus a
very convenient quantity to test the algorithms with.  The algorithms
presented in this paper can, however, be used for the same calculations
as other VBQMC implementations~(see e.g.  \cite{beach_formal_2006}).

In section~\ref{sec_tree_algorithm} we derive the general equations
governing the movement of the worm. Section~\ref{sec_impl} contains a
description of the specific implementation corresponding to the two
choices of parameter solutions we have studied. The \emph{bouncing worm}
is detailed in section~\ref{subsec_bouncing_worm} while the \emph{driven
worm} algorithm is described in section~\ref{subsec_driven_worm}.  The
algorithms are then compared in section~\ref{sec_comp}.  We present our
conclusions in section~\ref{sec_conclusion}.

\section{Tree Algorithms}
\label{sec_tree_algorithm}
We now turn to a discussion of the general framework for the algorithms
we have investigated.  We begin by deriving the equations governing
their behavior in a general way.

Let us take the Hamiltonian to have $N_B$ terms.  We now imagine a tree
where each node indicates the decision to chose one of the $N_B$ bond
operators composing the string (see Fig.~\ref{fig_tree}). Each branch of
the tree corresponds to one of the $N_B$ bond operators.  A given
operator string then corresponds to selecting a path in the tree.
Consider 2 such paths \textsf{\textit{S}} and \textsf{\textit{T}} that
are identical for the part of the operator string first applied to the
trial-state.   The last 3 operators, however, differ. This leads to
different weights, which we denote with $s_i$ and $t_i$.  

As it is done in most Monte Carlo methods, we set out to construct a
Markov-chain. Here it is a chain of different strings. If the
probabilities to go from one string to the next have detailed balance,
the Markov-chain contains the strings with the desired probability.  For
detailed balance,  the probabilities for starting from operator string
\textsf{\textit{S}} and going to operator string \textit{\textsf{T}} and
reverse have to satisfy
\begin{equation}
  \frac{P(\textit{\textsf{S}} \to \textit{\textsf{T}})}
  {P(\textit{\textsf{T}} \to \textsf{\textit{S}})}
  =
  \frac{t_3t_4t_5}{s_3s_4s_5}\;.
  \label{eq_mc}
\end{equation}
We can achieve this ratio of probabilities by imagining a worm
(tree-worm) working its way up the tree to the point $p$ where it turns
around and then working its way down again. 

Let us call the valence-bond covering of the trial-state $C_0$. Up to
numerical factors, the application of an operator string
\textit{\textsf{S}} of length $n$ will yield a new valence-bond covering
$\textit{\textsf{S}} \, C_0\propto C_n$.  The worm is started by
removing the last applied bond operator and considering the resulting
covering $C_{n-1}$. A decision now has to be made if the worm is to
continue "up" the tree by removing more bond operators from the string
or if it should instead go "down" the tree by adding a new bond operator
to the string.  At each node in the tree the decision to continue up or
turn around is made according to a set of \emph{conditional
probabilities} $P(\mathrm{up}|s)$ and $P(t|s)$. Here, $P(\mathrm{up}|s)$
denotes the probability for going up after coming from a bond operator
that carried weight $s$ and $P(t|s)$ is the probability for turning
around by applying a bond operator of weight $t$ coming from an operator
with weight $s$.  Likewise, $P(s|\mathrm{up})$ denotes the probability
of choosing an operator with weight $s$ given that the worm is coming
from further up the tree.  With these conditional probabilities the
left-hand side of Eq.~(\ref{eq_mc}) can be written as
\begin{widetext}
  \begin{equation}
    \frac{P(\textit{\textsf{S}} \to \textit{\textsf{T}})}
    {P(\textit{\textsf{T}} \to \textsf{\textit{S}})}  
    =
    \frac{P(t_5|\mathrm{up})P(t_4|\mathrm{up})P(t_3|s_3)P(\mathrm{up}|s_4)P(\mathrm{up}|s_5)}
    {P(\mathrm{up}|t_5)P(\mathrm{up}|t_4)P(s_3|t_3)P(s_4|\mathrm{up})P(s_5|\mathrm{up})}
    \;.
    \label{eq_detailed_balance}
  \end{equation}
\end{widetext}
Clearly, Eq.~(\ref{eq_detailed_balance}) is satisfied if we choose
\begin{equation}
  \frac{P(\mathrm{up}|s)}{P(s|\mathrm{up})}=\frac{c}{s}
  \hspace{2em}
  \textrm{and} 
  \hspace{2em}
  \frac{P(t|s)}{P(s|t)}=\frac{t}{s} \;,
\label{eq_ratio}
\end{equation}
where $c$ is an additional free parameter included for later
optimization of the probabilities.  If we can choose conditional
probabilities with these properties, we can go between different
operator strings always accepting the new string. The rejection
probability is then zero.  This is a very desirable property of any
Monte Carlo Algorithm since it indicates that the algorithm is sampling.
We mostly focus on so called \emph{zero bounce} algorithms for which if 
the worm turns around the probability for replacing a bond operator with
the same operator is zero. Then  the two operator strings $S$ and $T$
are always different. This means that
\begin{equation}
  P(s|s)=0 \,.
\end{equation}

Quite generally, it is  easy to find \emph{many} solutions to the
equations (\ref{eq_ratio}) leading to many Monte Carlo algorithms which
can be tuned for efficiency.  

We now focus on $S$\hspace{0pt}$=$\hspace{0pt}$1/2$-Heisenberg models
defined on bi-partite lattices.  As has been described above, for these
models only 2 weights can occur: $1,1/2$. The two weights correspond to
the two different actions the bond-operators can have on the state.  It
is $1$ if the operator acts on two sites that are in a valence-bond.
The state is not altered under the action of such an operator. We call
such operators diagonal. The weight is $1/2$ if the operator acts on two
sites that are \emph{not} in a valence-bond. After the action of the
operator the two sites are connected by a bond as well as the sites they
were connected to. We call such operators non-diagonal. 

If a decision has to be made at the node  at position $m$, the
conditional probabilities depend on how many of the $N_B$ bond-operators
will yield a weight of 1 (are diagonal) or $\frac{1}{2}$ (are
non-diagonal) when applied to the present covering $C_{m-1}$. We shall
denote these numbers by $N_1$ and $N_\frac{1}{2}$ respectively. When the
worm is started $N_1$ and $N_\frac{1}{2}$ therefore have to be
calculated for $C_{n-1}$, if they are not already known from an earlier
update. It is thus sensible to store $N_1$ or $N_\frac{1}{2}$ at all
nodes.  $N_1$ can only be zero at the node furthest up the tree (the
root) and only if the trial-state is chosen to not contain any diagonal
bonds. In Fig.~\ref{fig_tree} it is the gray node on the very left.
$N_{\frac{1}{2}}$ cannot be smaller than $N_B/2$. 

We can now write down an $(N_B+1)\times (N_B+1)$ matrix $M$ of
conditional probabilities for each node of the tree. The $j$'th column
of the matrix describes the probability for going in any of the $N_B+1$
directions when coming from the direction $j$. For clarity we order the
rows and columns such that the first $N_{\frac{1}{2}}$ correspond to the
non-diagonal operators and the next $N_1$ to the diagonal operators. The
last column contains the probabilities for going down the tree when
coming from above; the last row the probabilities for going up the tree
when coming from below. The remaining part of the matrix describe the
probabilities for replacing one operator with another when the worm
turns from going up to going down.
The matrix $M$ has the form
\begin{widetext}
  \begin{eqnarray}
    \overbrace{\rule{6.7cm}{0pt}}^{N_{\frac{1}{2}}}\overbrace{\rule{5.15cm}{0pt}}^{N_1}\overbrace{\rule{1.8cm}{0pt}}^\mathrm{up}\rule{0.4cm}{0pt}\nonumber\\
    M =    
    \left(
    \begin{array}{ccccccccccccc}
      P(\frac{1}{2}|\frac{1}{2}) & P(\frac{1}{2}'|\frac{1}{2}) & P(\frac{1}{2}'|\frac{1}{2}) & \cdots  & P(\frac{1}{2}'|\frac{1}{2})&
      \vline & P(\frac{1}{2}|1) & P(\frac{1}{2}|1) &\cdots & P(\frac{1}{2}|1) & \vline&  P(\frac{1}{2}|\mathrm{up})\\

      P(\frac{1}{2}'|\frac{1}{2}) & P(\frac{1}{2}|\frac{1}{2}) & P(\frac{1}{2}'|\frac{1}{2}) & \cdots  &\cdot &
      \vline& \cdot & \cdot & \cdots & \cdot & \vline&\cdot \\

      \cdot & \cdot & \cdot &        & \cdot & \vline &\cdot& 
      \cdot &       & \cdot & \vline & \cdot \\

      \cdot & \cdot & \cdot &        & \cdot & \vline &\cdot& 
      \cdot &       & \cdot & \vline & \cdot \\

      P(\frac{1}{2}'|\frac{1}{2}) &P(\frac{1}{2}'|\frac{1}{2})&P(\frac{1}{2}'|\frac{1}{2}) & \cdots & P(\frac{1}{2}|\frac{1}{2})&
      \vline & P(\frac{1}{2}|1) & P(\frac{1}{2}|1) & \cdots & P(\frac{1}{2}|1) & \vline & P(\frac{1}{2}|\mathrm{up}) \\

      \hline

      P(1|\frac{1}{2}) & P(1|\frac{1}{2}) & P(1|\frac{1}{2}) & \cdots & P(1|\frac{1}{2})&
      \vline&P(1|1) & P(1'|1) & \cdots & P(1'|1) & \vline & P(1|\mathrm{up}) \\

      \cdot &  \cdot & \cdot  &        & P(1|\frac{1}{2}) & 
      \vline & P(1'|1) & P(1|1) & \cdots & \cdot & \vline &\cdot \\

      \cdot &  \cdot & \cdot  &        & \cdot & 
      \vline & \cdot & \cdot &      & \cdot &   \vline&\cdot \\

      \cdot &   \cdot             & \cdot &         & \cdot & 
      \vline &\cdot& \cdot &      & \cdot &   \vline&\cdot \\

      P(1|\frac{1}{2}) & P(1|\frac{1}{2}) & P(1|\frac{1}{2}) &            \cdots  & P(1|\frac{1}{2})&
      \vline&P(1'|1) & P(1'|1) &\cdots & P(1|1) & \vline&P(1|\mathrm{up}) \\

      \hline

      P(\mathrm{up}|\frac{1}{2}) & P(\mathrm{up}|\frac{1}{2})  & P(\mathrm{up}|\frac{1}{2})  &            \cdots  & P(\mathrm{up}|\frac{1}{2})&
      \vline &P(\mathrm{up}|1) & P(\mathrm{up}|1) &\cdots & P(\mathrm{up}|1) & \vline& P(\mathrm{up}|\mathrm{up}) \\
    \end{array}
    \right) \; , 
    \nonumber
  \end{eqnarray}
\end{widetext}
where $P(s'|s)$ refers to the conditional probability of coming from an
operator with weight $s$ and going to a \emph{different} operator with
the same weight.  As mentioned above, $P(\mathrm{up}|s)$ denotes the
probability for going up coming from an operator with weight $s$ and
$P(t|s)$ is the probability for turning around by choosing a bond
operator of weight $t$ coming from an operator with weight $s$.
Likewise, $P(s|\mathrm{up})$ denotes the probability of choosing an
operator with weight $s$ coming from further up the tree.

To shorten the notation we introduce the short-hand
\begin{equation}
  x=P({1}/{2}'|{1}/{2}),\ \ y=P(1/2|1),\ \ z=P(1'|1)\;.
\end{equation}
Furthermore we define the `bounce' probabilities
\begin{equation}
  b_{\frac{1}{2}}=P(1/2|1/2),\ \ b_1=P(1|1),\ \
  b_u=P(\mathrm{up}|\mathrm{up})\;.
\end{equation}
Here it is implied that the probabilities are for going from one
operator to the \emph{same} operator.  Finally we also need to define the
branching probabilities
\begin{equation}
  u=P(1/2|\mathrm{up}), \ \ w=P(1|\mathrm{up}) \;,
\end{equation}
from which it follows (using Eq.~(\ref{eq_ratio})) that:
\begin{equation}
  2cu=P(\mathrm{up}|1/2), \ \ cw=P(\mathrm{up}|1) \;.
\end{equation}

The matrix $M$ is then given by 
\begin{equation}
  \hspace{-20pt}
  M =
  \left(
  \begin{matrix} 
    b_{\frac{1}{2}} & x & x & \cdots  & x & y & y &\cdots & y &   u\\

    x & b_{\frac{1}{2}} & x & \cdots  & x & y & y & \cdots & y & u \\

    \cdot & \cdot & \cdot &         & \cdot & 
    \cdot & \cdot &      & \cdot &   \cdot \\

    \cdot & \cdot & \cdot &         & \cdot & 
    \cdot & \cdot &      & \cdot &   \cdot \\

    x     & \cdot & \cdot & \cdots   &               b_{\frac{1}{2}} & y & y & \cdots & y & u \\

    2y & 2y & 2y &            \cdots  & 2y&b_1 & z &\cdots & z & w \\

    \cdot &  \cdot & \cdot  &              & 2y &z  & b_1 &\cdots & \cdot & \cdot \\

    \cdot &   \cdot             & \cdot &         & \cdot & 
    \cdot & \cdot &      & \cdot &   \cdot \\

    \cdot &   \cdot             & \cdot &         & \cdot & 
    \cdot & \cdot &      & \cdot &   \cdot \\

    2y & 2y & 2y &            \cdots  & 2y&z & z &\cdots & b_1 & w \\
    2cu & 2cu & 2cu &            \cdots  & 2cu&cw & cw &\cdots & cw & b_u \\
  \end{matrix}
  \right)
  \;.
  \label{eq_defmatrix}
\end{equation}
The requirement that this matrix be stochastic (i.e.\ some branch is
chosen with probability one) means that the entries in each column have
to sum to 1.  This leads to the set of equations
\begin{eqnarray}
  1&=&N_{\frac{1}{2}}u+N_1w+b_u\nonumber\\
  1&=&N_{\frac{1}{2}}y+(N_1-1)z+cw+b_1\nonumber\\
  1&=&(N_{\frac{1}{2}}-1)x+N_12y+2cu+b_{\frac{1}{2}}\;.
  \label{eq_cond}
\end{eqnarray}
These simple equations are the central equations governing the behavior
of the algorithms.  To find an algorithm, we need to solve these 3
equations with the constraints that $0\le
x,y,z,b_{\frac{1}{2}},b_1,b_u,u,w\le 1$;  a straight forward problem. 

At the root, the equations are modified slightly: since it is not
possible to go further up the tree, $2cu$, $ cw $, $ b_u$ are not
meaningful and can be set to zero.  For convenience we set $b_1 = x$ and
$b_2 = z$ at the root. This allows one to just choose diagonal operators
twice as often as non-diagonal operators.  Since the number of diagonal
operators does not change at the root, a table generated at the beginning
of the calculation suffices to perform this task.

It can be very useful to choose different $c$'s at different nodes.
Then, calculating the probabilities to choose operators according to the
rules introduced in this section will not lead to an algorithm with
detailed balance, because $c_i$ from different strings will not cancel
in Eq.~(\ref{eq_detailed_balance}). It is necessary to work with an
acceptance probability. 
We find
\begin{equation}
    \frac{P(\textit{\textsf{S}} \to \textit{\textsf{T}})}
    {P(\textit{\textsf{T}} \to \textsf{\textit{S}})}  
    =
  \frac{t_3t_4t_5}{s_3s_4s_5}
  \frac{c^\textit{\textsf{S}}_4c^\textit{\textsf{S}}_5}{c^\textit{\textsf{T}}_4c^\textit{\textsf{T}}_5}
    \frac{P_{\mathrm{acc}}(\textit{\textsf{S}} \to \textit{\textsf{T}})}
    {P_{\mathrm{acc}}(\textit{\textsf{T}} \to \textsf{\textit{S}})} \;.
  \label{eq_mc2}
\end{equation}
Here $c^\textit{\textsf{S}}_i$ and $c^\textit{\textsf{T}}_i$ denote $c$
at the different nodes in the strings $\textit{\textsf{S}}$ and
$\textit{\textsf{T}}$, respectively. To validate the algorithm, we must
therefore introduce an acceptance probability that must cancel the
factor
$(c^\textit{\textsf{S}}_4c^\textit{\textsf{S}}_5)/
(c^\textit{\textsf{T}}_4c^\textit{\textsf{T}}_5)$.
This can be achieved by choosing
\begin{equation}
  P_{acc}(\textit{\textsf{S}}\to
  \textit{\textsf{T}})
  =
  \min\Big(1,\frac{c^\textit{\textsf{T}}_4c^\textit{\textsf{T}}_5}
  {c^\textit{\textsf{S}}_4c^\textit{\textsf{S}}_5}\Big) \;,
  \label{eq_accprob}
\end{equation}
meaning that when a new string is generated through a worm move it is
accepted with this probability. 

Since we always start from the bottom of the tree (the last operator
applied), the worm algorithms presented in this paper always change a
block of consecutive branches at the end of the string. This is
favorable to changes across the whole string because changes far up the
string might be undone by changes closer to the end of the
string~\cite{sandvik_monte_2007}. In this way the most important part of
the string is updated most substantially.

It is also important to note that the algorithm will conserve certain
topological numbers. For instance, for a two-dimensional system $S=1/2$
Heisenberg model the number of valence bond crossing a cut in the $x$-
or $y$-direction is either odd or even. Hence, the initial covering,
$C_0$ is characterized by these 2 parities. It is easy to see that the
application of $P$ to any covering can not change these parities and
they are therefore preserved under the projection.

\section{Implementations of tree worm algorithms}
\label{sec_impl}
As is explained in the last section, many different algorithms can be
found because many different solutions to the equations~(\ref{eq_cond})
exist.  

In this section we present two  different algorithms.  One pure
worm algorithm where every update is accepted (the \emph{bouncing worm}
algorithm) and an algorithm that allows for control over how far in the
tree updates are attempted (the \emph{driven worm} algorithm). 
To test and compare the different algorithms, we calculate the
ground-state energy of the antiferromagnetic Heisenberg chain.

The N\'{e}el-state $|\textrm{N\'eel}\rangle$ has equal overlap with all
valence-bond states.  This can be used to very directly estimate the
ground-state energy, 
$E_0$~\cite{sandvik_ground_2005,*beach_formal_2006,*sandvik_monte_2007}:
\begin{align}
  E_0 & =  \frac{\langle \textrm{N\'eel}| H |\Psi_0\rangle}
  {\langle \textrm{N\'eel}|  \Psi_0 \rangle}
  \nonumber \\
   &= 
  \lim_{n\to \infty} \frac{\langle \textrm{N\'eel}| H P^n |C_0\rangle}
  {\langle \textrm{N\'eel}|  P^n |C_0\rangle}
  \nonumber\\
  & = 
  \lim_{n\to \infty}
  \sum_{a=1}^{N_B^n} 
  \frac{\langle \textrm{N\'eel}| H S_a|C_0\rangle}
  {\sum_{a=1}^{N_B^n} \langle \textrm{N\'eel}|  S_a|C_0\rangle}
  \nonumber \\
  & = 
  \lim_{n\to \infty}
  \sum_{a=1}^{N_B^n} 
  \frac{W_a}
  {\sum_{a=1}^{N_B^n} W_a } 
  \frac{\langle \textrm{N\'eel}| H |C_a\rangle}
  {\langle \textrm{N\'eel}|  C_a\rangle}.
  \label{eq_est1}
\end{align}
If we take $E_aC_b=HC_a$ and assume that the Monte Carlo sampling will
visit strings according to their weight $W_a$, then for a Monte Carlo
sequence of length $N$ of independent strings we find:
\begin{equation}
  E_0=\frac{1}{N}
  \sum_{a=1}^N E_a,
  \label{eq_estimator}
\end{equation}
where again we have used the fact that $\langle \textrm{N\'eel}| C\rangle$ is independent of
the covering $C$.  

To analyze the correlation-properties of the worm algorithms we use the
energy-autocorrelation-time, which we take to be the number of updates
it takes the energy-autocorrelation-function 
\begin{align}
  A_E(t) =
  \frac{\langle E_i E_{i+t}\rangle - \langle E \rangle ^2}
  {\langle E^2 \rangle - \langle E \rangle ^2} 
  \label{eq_def_autocorr}
\end{align}
to decay to 0.1. The results of all update-attempts enter the
calculation of the expectation-values. The shorter the
autocorrelation-time is, the fewer steps have to be done between
consecutive measurements.

If not stated otherwise an operator-string of 20,000 operators was used
for calculations with worm algorithms.

\subsection{The bouncing worm algorithm}
\label{subsec_bouncing_worm}
\begin{figure}[t]
  \includegraphics[width=.85\linewidth]{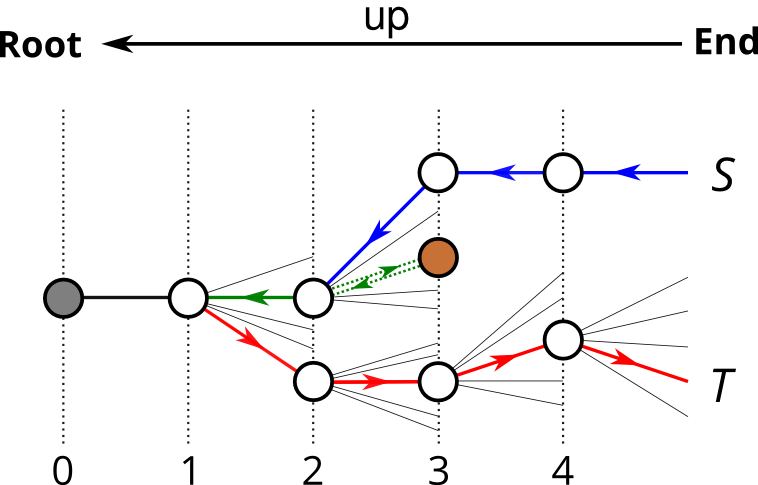}
  \caption{A possible path that contains one bounce and connects the
    string \textit{\textsf{S}} and the string \textit{\textsf{T}}. The
    worm first goes up to the node $2$ where it turns to go down to node
    3.  The worm bounces back and goes all the way to node 1. Then the
    worm turns around and does not bounce again.
  }
  \label{fig_treebouncing}
\end{figure}
The first algorithm we discuss is the bouncing worm algorithm.  Only a few
of the variables that appear in the equations~(\ref{eq_cond}) are chosen
to be non-zero. We choose to set:
\begin{equation}
x=y=0, \ \ b_1=b_{1/2}=0,
\end{equation}
while $z=P(1'|1)\neq 0$ as is $u,w$. We leave $b_u$ as a parameter that
can be zero or non-zero allowing for tuning of the algorithm.  With this
choice, when the worm is moving up the tree the only possibility for it
to turn around is by opting to replace one diagonal operator with
another diagonal operator.  The $c_i$ are chosen to be the same at all
nodes: $c_i = c$. 

The equations for the non-zero parameters are then
\begin{eqnarray}
  u&=&\frac{1}{2c}\nonumber\\
  w&=&\frac{1}{N_1}\Big(1 - \frac{N_\frac{1}{2}}{2c} - b_u \Big)\nonumber\\
  z&=&\frac{1 - cw}{N_1-1}\;. 
  \label{eq_algoD_weights}
\end{eqnarray}
The requirements that $z,w>0$ imply that
\begin{equation} 
  \frac{N_\frac{1}{2}}{2 (1- b_u)}\leq c\leq\frac{N_B+N_1}{2(1 - b_u)} \;.
  \label{eq_algoD_ccond}
\end{equation}
To satisfy Eq.~(\ref{eq_algoD_ccond}) with node-independent $c$, we set 
\begin{align}
  c = \frac{N_B}{2 (1-b_u)}  \;.
  \label{eq_bounce_c}
\end{align}

With this choice of parameters, we find the
probability to go up the tree if the worm is at a node with a non-diagonal
operator to be
\begin{equation}
P(\mathrm{up}|1/2)=2cu = 1 \,,
\end{equation}
for \emph{any} $b_u$.  Likewise, if the worm is at a node with a
diagonal operator the probability to go up is given by 
\begin{equation}
P(\mathrm{up}|1)=
c w = 1 / 2 \,,
\end{equation}
independent of $b_u$. The probability for going up the tree is therefore
independent of $b_u$.

We define the penetration depth (p.-depth), which we denote by $r$, as
the maximal height that the worm reaches.  The actual length of the worm
is denoted by $l$ and with $b_u=0$ we find $l=2r$. The penetration depth
$r$ will determine how much the operator-string is changed. Obviously,
it is desirable to have the worm reach as far up the tree as possible.
It is possible to force the worm farther up the tree by having it bounce
back to going up after it has turned to go down (see
Fig.~\ref{fig_treebouncing}).  In that case, the actual {\it length} of
the worm, $l$, will then be substantially different from twice the
penetration depth since the
worm can turn many times, a point we shall return to later.  Such
bounces occurs with a likelihood of $b_u$ which was left as a free
parameter and can now be used as a tuning parameter.

The algorithm is straight forward to implement and the acceptance
probability for a worm update is 1. The move is \emph{always} accepted.
Specific details of an implementation of the bouncing worm update can be
found in appendix~\ref{subsec_bouncing_worm_pse}. 

We begin by discussing the case of $b_u=0$. In this case the worm first
moves up the operator string, turns around once and then proceeds down to
the bottom of the tree. It does not go back up the operator string since
$b_u=0$.  In order to measure the performance of the algorithm we 
did calculations on an antiferromagnetic Heisenberg chain with 50 sites
using an operator string of length $100,000$. As can be seen in
table~\ref{tab_bounce_slowdown}, this leads to a rather small mean
penetration-depth (p.-depth) of about 5.   The maximal penetration-depth
of 50 is substantially larger.  Both these numbers are, however,
substantially smaller than the length of the operator string ($100,000$)
and it appears that the algorithm with $b_u$ is not very effective.
\begin{table}[htb]
  \begin{tabular}{c@{\hspace{.5cm}}c@{\hspace{.5cm}}c@{\hspace{.5cm}}c}
    \toprule[.08em]
    $b_u$ & mean p.-depth & max p.-depth & slowdown \\
    \midrule[.05em]
    0.0000    &  4.561(4)   & 50  & 1 \\
    0.2500    &  7.38(1)  & 305 & 1.7 \\ 
    0.2750    &  10.44(3)  & 1,465 & 9.7 \\
    0.2789    &  15.64(9)  & 41,010 & 316.7 \\
    0.2790    &  19.7(5)  & $>$100,000 & $<$5,535.7 \\
    \bottomrule[.08em]
  \end{tabular}
  \caption{Data for several runs at different $b_u$. At $b_u\approx
    0.25$ increasing the bounce-probability  starts to significantly
    slow down the algorithm.  The last column contains the run-times
    divided by the runtime for $b_u$~$=$~$0$.  The data were generated
    with an operator-string of 100,000 operators. The maximal
    penetration and the expected slowdown could thus not be resolved for
    $b_u$ $=$ $0.2790$. We used $10^6$ measurements and a chain with 50
    sites. 
  }
  \label{tab_bounce_slowdown}
\end{table}

We now turn to the case $b_u\neq 0$. In this case the worm can now
switch directions many times during construction (see
Fig.~\ref{fig_treebouncing}).  The results for the mean and maximal
penetration-depth are also listed in table~\ref{tab_bounce_slowdown}.
As $b_u$ is increased from zero, the maximal penetration-depth first
increases very slowly until about $b_u=0.25$. It then grows dramatically
and, not surprisingly, reaches the length of the operator string. This
occurs at $b_u\approx0.2790$.  At the same time the mean
penetration-depth only increases by a factor of roughly 4, from 5 to
about 20.  For bounce-probabilities bigger than $b_u \approx 1 / 4$ the
program is slowed down significantly compared to the algorithm with
$b_u=0$ as indicated in the last column in
table~\ref{tab_bounce_slowdown}.  Thus, even though a large
penetration-depth is desirable the computational cost can become so big
that increasing $b_u$ might not be worthwhile.

\begin{figure}[htb]
  \includegraphics[]{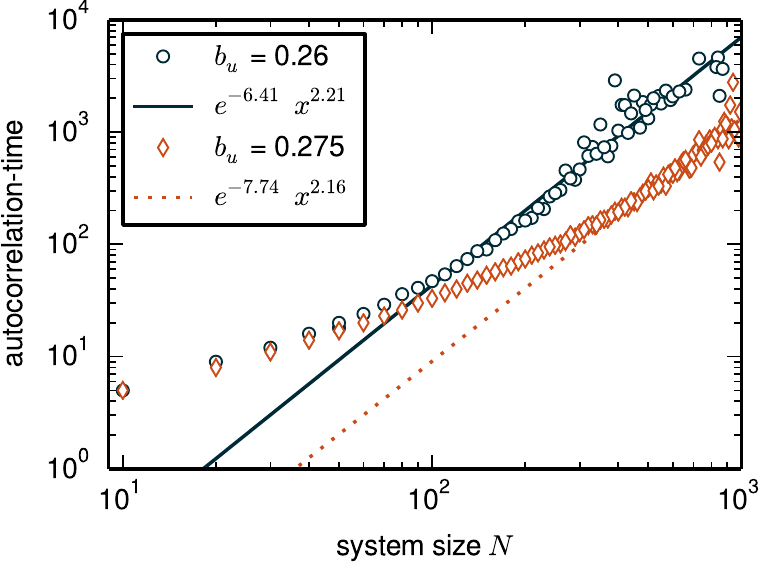}
  \caption{The autocorrelation-time of the energy as a function of
    system size $N$.  The lines indicate tentative power-law fits to the
    data at large $N$ with a power of $2.21$ for $b_u=0.26$ and $2.16$
    for $b_u=0.275$. Operator-strings of 20,000 operators were used.
  }
  \label{fig_bounce_autocor}
\end{figure}

In contrast to the maximal penetration-depth, the mean penetration-depth
grows very slowly for the values of $b_u$ we have been able to study.
For computations of reasonable computational cost it never reaches the
size of the system and thus also not the length of the operator-string
which has to be chosen to be several times the size of the system.  That
only a small part of the string is updated regularly is directly
reflected in the energy-autocorrelation-time (see
Fig.~\ref{fig_bounce_autocor}).  The number of bonds that can change in
one update of the worm-calculations is twice the penetration-depth.
Typical updates never reach far into the operator-string. Thus, the
bigger the system is, the less it is perturbed by the update and the
more correlated are the energies measured after consecutive updates.  
As shown in Fig.~\ref{fig_bounce_autocor}, increasing
$b_u$ decreases the autocorrelation-time. However, for large system
sizes the overall  scaling of the autocorrelation-time with the system
size appears independent of $b_u$.  At $b_u=0.26$ we find a power-law
with an exponent of $2.21$ while a slightly larger $b_u=0.275$ yields a
power of $2.16$.

Even though the mean penetration depth remains small, one can still obtain
high quality results. In particular, it is \emph{not} necessary for the mean penetration depth to reach
a value close to the length of the operator-string (the projection power) in order to get reliable results.
Since the maximal penetration-depth is substantially larger than the
mean, the operator-string is often updated deeper than the mean
penetration-depth.  Hence,  the mean penetration-depth can be much
smaller than the length of the operator-string has to be for otherwise
equivalent calculations with conventional VBQMC\@.
We discuss this effect in more detail in section~\ref{sec_comp}.
\begin{figure}[htb]
  \includegraphics[]{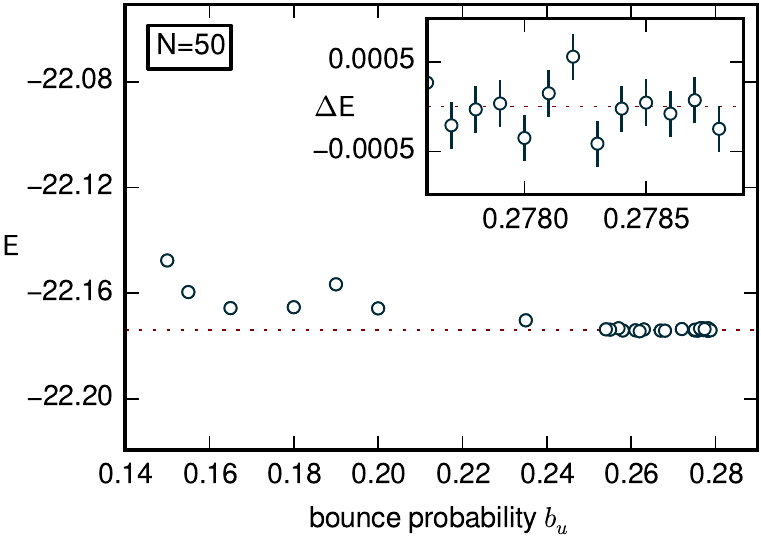}
  \caption{With a bigger probability to bounce $b_u$, the bouncing worm
    algorithm yields a better approximation of the ground-state energy.
    The ground-state energy was calculated using the Bethe-ansatz and is
    indicated by a dotted line.  Since some operators in the string are
    never updated, calculations with different $b_u$ were effectively
    done with different trial-states. The calculation was done for a
    chain with 50 sites.  Operator-strings of 20,000 operators were
    used.   
  }
  \label{fig_bounceworm_proofofworking}
\end{figure}

As one increases the probability to bounce up the tree a longer part of
the string actually partakes in the projection. Thus, also the quality
of the projection is better (see
Fig.~\ref{fig_bounceworm_proofofworking}).  If operators are never
updated, they do not contribute to the projection; they do however
modify the trial-state used in the projection. This leads to the
irregularly scattered pattern the data shows for small $b_u$. In this
sense one can think of rare updates that go high up the tree as
effectively changing the trial-state and the whole calculation as an
averaging over these trial-states.

As is evident from Fig.~\ref{fig_bounceworm_proofofworking}, the
bouncing worm algorithm yields good results for $b_u>0.25$. It is an
attractively simple algorithm with zero bounce probability and a
probability of 1 for accepting a new string.  The autocorrelation-time
can be  reduced by increasing $b_u$ whereas the overall power of the
growth at large system sizes appears independent of $b_u$.  However, the
increased computational cost associated with increasing $b_u$ is
considerable and we have therefore investigated another parameter choice
leading to a different algorithm, the driven worm algorithm.  We now
turn to a discussion of this algorithm.

\subsection{The driven worm algorithm} 
\label{subsec_driven_worm}
Clearly, it is desirable to have all updates result in a substantial
change of the operator-string. Then, fewer updates have to be performed.
For the problem at hand, this means that we need the worm to go far up
the tree as often as possible without increasing the computational cost
too much. Direct control over the associated probability would be
very convenient.  We achieve this by setting the probabilities to go up
the tree to be
\begin{equation}
  2cu = cw = \alpha \,.
  \label{eq_alphadefinition}
\end{equation}
The value of $\alpha$ is the probability to, at each node, decide to go
up the tree.  Since $u$ and $w$ depend on $N_1$ and $N_2$, this is only
possible by allowing $c$ to vary with the node. As explained at the end
of section~\ref{sec_tree_algorithm}, the acceptance step of
Eq.~(\ref{eq_accprob}) thus has to be introduced. Updated strings may be
rejected.

We set all bounce-probabilities to be zero, $b_1=b_{1/2}=b_u=0$.  Hence,
the worm will move up the tree and then turn around once.  To get a
working algorithm, we have to find solutions to the
equations~(\ref{eq_cond}) which will determine the
transition-probabilities (see Eq.~(\ref{eq_defmatrix})).  If
$N_{\frac{1}{2}}, N_1 > 1$ the solutions to equations~(\ref{eq_cond})
are given by:
\begin{align}
  \label{eq_algoC_weights}
  x =&\frac{1}{N_{\frac{1}{2}} ( N_{\frac{1}{2}} - 1)} \Big[ 2 N_1
    (N_1  -1) \ z \nonumber \\ 
  &+ (1-\alpha) (N_{\frac{1}{2}} - 2 N_1)\Big] \nonumber\;, \\
  y =& \frac{1}{N_{\frac{1}{2}}} \Big[(1-N_1) z + 1 -\alpha
  \Big]\nonumber \;,\\
  u=&1 / (  N_{\frac{1}{2}} + 2 N_1)\nonumber \;, \\
  w = & 2 u \;, 
\end{align}
where
\begin{eqnarray}
  \label{eq_algoC_zcond}
  \frac{1-\alpha}{N_1 -1} \Big[ 1- \frac{N_{\frac{1}{2}}}{2
  N_1}\Big]\leq z  \leq \frac{1-\alpha}{N_1 - 1} \;. 
\end{eqnarray}
If $N_1 \neq 1$, we set
\begin{equation}
  z = \frac{1-\alpha}{N_1 -1} \Big[ 1-\frac{1}{2} \frac{N_{\frac{1}{2}}}{2
  N_1}\Big]  \  ,
\end{equation}
if it results in $z>0$ or 
\begin{equation}
  z  =  \frac{1}{2}\frac{1-\alpha}{N_1 - 1} \ \\ 
\end{equation} 
otherwise.  In this way Eq.~(\ref{eq_algoC_zcond}) is always satisfied
and $z\geq0$.  If $N_1 = 1$ we set $z=0$.  Finally, we note that the
worm update in this case has to be accepted/rejected according to the
probability Eq.~(\ref{eq_accprob}).  Specific details of an
implementation of this driven worm algorithm can be found in
appendix~\ref{subsec_driven_worm_pse}.

How far up the tree updates are attempted can in this case easily be
calculated.  The probability for the worm to 
have length $l$
and turn around after going up $r =l/2$ nodes is given by $P(r) =
\alpha^{r}(1-\alpha)$.  The expectation-value of $r$ is given by
\begin{align}
  \langle r \rangle = \frac{1}{1-\alpha}  \;.
  \label{eq_algoC_mean_p_depth}
\end{align}

The probability distribution for the worm to penetrate the tree $r$
nodes deep during a computation of $m$ updates, is given by 
\begin{align}
  P_{m,\alpha}(r_{\textrm{max}}) = & 
  \underbrace{%
    \Big(1 - (1 - \alpha)  \sum_{q=r}^\infty \alpha^q \Big)^m
  }_{\substack{%
    \text{probability that in $m$ attempts \emph{no} worm} \\
    \text{turns at a node with $r>r_{\textrm{max} }$}
  }}  \nonumber \\
  & - \underbrace{%
    \Big( (1-\alpha) \sum_{q=1}^{r-1} \alpha^q \Big)^m
  }_{\substack{%
    \text{probability that in $m$ attempts \emph{all} worms} \\ 
    \text{turn at a node with $r<r_{\textrm{max}}$} 
  }}  \nonumber \\[10pt]
  =& (1 - \alpha^{r})^m - (1-\alpha^{r-1})^m \;.
  \label{eq_algoC_max_p_depth}
\end{align}

\begin{figure}[htb]
  \includegraphics[]{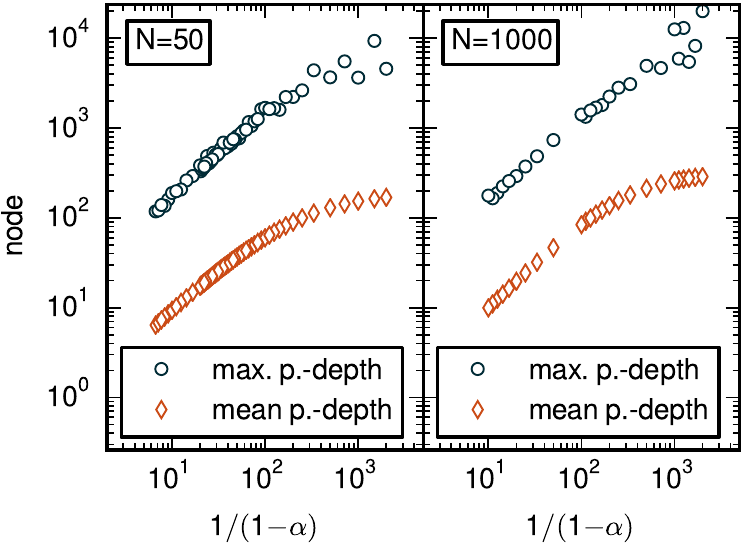}
  \caption{The mean and the maximal penetration-depths grow as $\alpha$
    approaches one. The mean penetration-depth is always smaller than
    $\langle r \rangle = 1/ ( 1- \alpha)$. This behavior is independent
    of the system size.  For the chain with 50 sites $\mathrm{3} \times
    \mathrm{10}^{\mathrm{8}}$ and for the chain with 1000 sites
    $\mathrm{10}^{\mathrm{7}}$ updates were performed.  
    Operator-strings of 20,000 operators were used.
  }
  \label{fig_driven_meann_maxn}
\end{figure}

How far up the tree is updated, is \emph{not} given by how far the worm
goes up the tree since the update might be rejected. The mean
penetration-depth is therefore not equal to $\langle r \rangle$. In
Fig.~\ref{fig_driven_meann_maxn} we show results for the mean and
maximal penetration-depth for two different system sizes, $N=50,1000$ as
a function of $1/(1-\alpha)$.  As expected, both the mean and maximal
penetration-depth increase monotonically with $1/(1-\alpha)$.

\begin{figure}[htb]
  \includegraphics[]{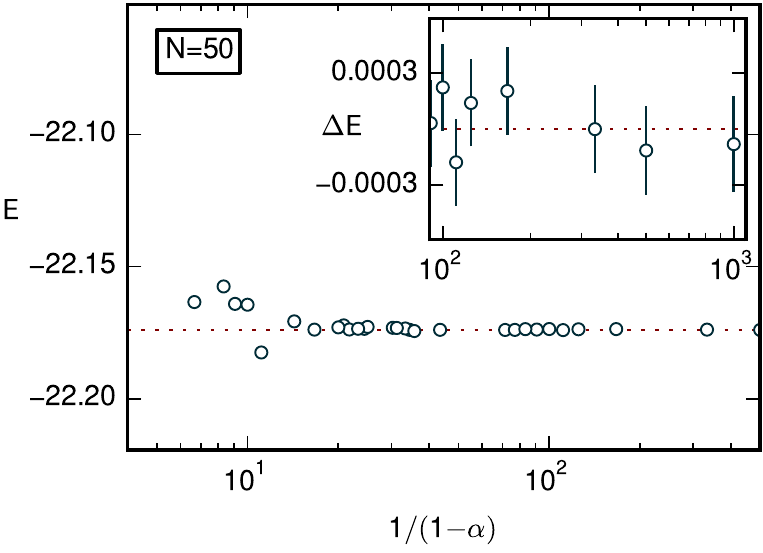}
  \caption{The bigger the penetration probability $\alpha$, the better
    the approximation of the ground-state energy given by the driven
    worm algorithm. The ground-state energy was calculated using the
    Bethe-ansatz. It is indicated by a dotted line. Since some operators
    in the string are never updated, calculations with different
    $\alpha$ were effectively done with different trial-states. The
    calculation was done for a chain with 50 sites.  
    Operator-strings of 20,000 operators were used. 
  }
  \label{fig_driven_conv}
\end{figure}

Another measure of the performance of the algorithm can be established
by simply looking at the calculated ground-state energy and its error.
This is done in Fig.~\ref{fig_driven_conv} where the ground-state energy
is shown as a function of $1/(1-\alpha)$.
Operators that are never updated, only change the effective trial-state
the ground-state is projected out of.  By forcing the worm further up
the tree, one can have a bigger part of the operator-string partake in
the projection (see Fig.~\ref{fig_driven_meann_maxn}). This leads to a
better approximation of the ground-state energy as can be seen in 
Fig.~\ref{fig_driven_conv}.

\begin{figure}[htb]
  \includegraphics[]{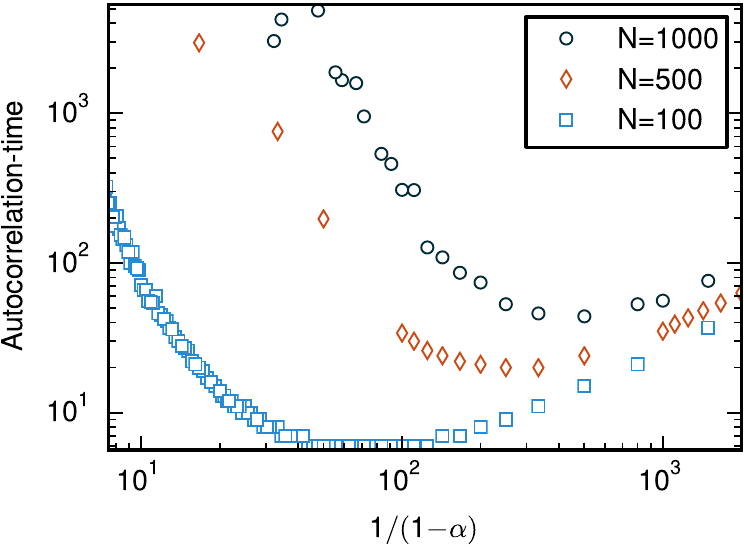}
  \caption{The autocorrelation-time decreases with $\langle r \rangle$,
    if typical updates are smaller than  the system size. Since bigger
    $\langle r \rangle$ means better projection, this implies that the
    autocorrelation-time \emph{decreases} as the quality of the
    projection is improved.  Operator-strings of 20,000 operators were
    used. 
  }
  \label{fig_driven_autocorti}
\end{figure}

For the driven worm algorithm we have also studied the behavior of the
autocorrelation-time of the energy. Our results are shown in
Fig.~\ref{fig_driven_autocorti} as a function of $1/(1-\alpha)$.  The
behavior is in this case not monotonic. At first the autocorrelation
time decreases, but then it starts to \emph{grow} at larger $1/(1-\alpha)$.

This can be understood in the following way: As long as $\langle r
\rangle$ is much smaller than the size of the system, the
autocorrelation-time \emph{decreases} with increasing $\alpha$. This 
follows naturally from the fact that increasing $1/(1-\alpha)$ will
increase $\langle r \rangle$ and therefore lead to larger and more
effective updates. This decreases the correlations between operator-strings.
The farther the worm travels up the string, the smaller is the
probability that an update is accepted (see
table~\ref{tab_acceptancerates}). For bigger $\alpha$, and thus also
$\langle r \rangle$, this effect dominates and the autocorrelation-time
grows. A characteristic minimum in the autocorrelation-time as
$1/(1-\alpha)$ is increased can then be identified as is clearly evident
in Fig.~\ref{fig_driven_autocorti}.  

\begin{table}
  \begin{tabular}{c@{\hspace{.5cm}}c}
    \toprule[.08em]
    $1/ (1-\alpha)$ & acceptance-rate \\
    \midrule[.05em]
    200 & 0.41 \\
    1000 & 0.13 \\
    5000 & 0.03 \\
    \bottomrule[.08em]
  \end{tabular}
  \caption{The acceptance rate drops when the string is updated
    more substantially. The calculations were done for a chain with 50
    sites.
  }
  \label{tab_acceptancerates}
\end{table}

\section{Comparison of algorithms}
\label{sec_comp}

In the following we compare worm-updates to simple conventional
VBQMC-updates as described for example in
reference~\cite{sandvik_monte_2007}. This means that for VBQMC  we
attempt to change 4 randomly selected operators during one update.  We
do not compare to loop-updates as introduced in
reference~\cite{sandvik_loop_2010}, since we anticipate the worm
algorithms to be of most with algorithms for which loop-updates are not
known although our current implementations of them are similar to
conventional VBQMC.

\begin{figure}[htb]
  \includegraphics[]{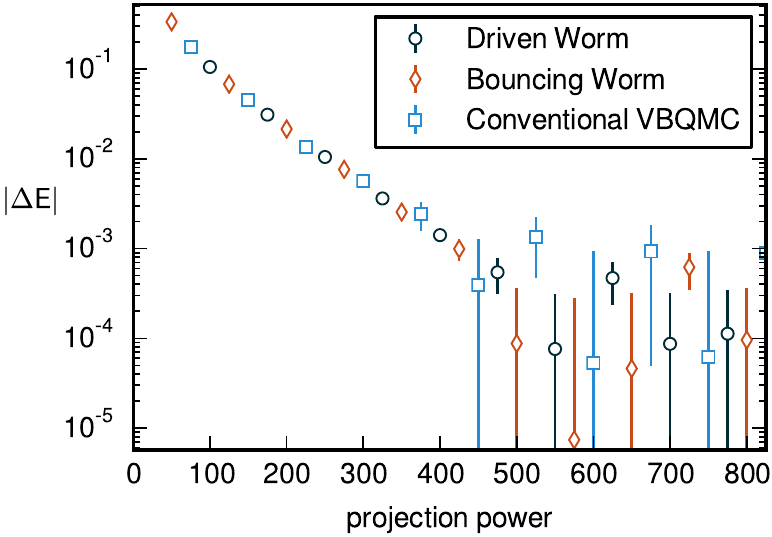}
  \caption{Upon increasing the quality of the projection by using longer
    operator strings, the estimate of the ground-state energy converges
    to the correct value in the same way for worm- and conventional
    VBQMC-algorithm as long as the string is penetrated sufficiently
    deeply. The driven worm algorithm was run with the probability to go
    up the tree $\alpha=0.995$, which corresponds to a mean
    penetration-depth of about $90$ and full penetration of the string.
    The bouncing worm algorithm  was run with a bounce probability
    $b_u=0.2675$, which corresponds to a mean penetration-depth of
    roughly  $8$ and full penetration of the string.  The calculation
    was done for a chain with 50 sites.  
  }
  \label{fig_comp_sivabo_driven_bounce}
\end{figure}

We first consider the convergence of the energy with the projection
power (the length of the operator strings). Our results for a 50 site
Heisenberg chain are shown in Fig.~\ref{fig_comp_sivabo_driven_bounce}.
It turns out that if the worm algorithms  penetrate the tree
sufficiently deeply, the results do not depend on the type of algorithm
in use.  In particular, the dependence of the results on the length of
the string is the same for \emph{all three} algorithms (see
Fig.~\ref{fig_comp_sivabo_driven_bounce}), just as one might have expected
since the power method underlies all three algorithms. 

When using the worm algorithms, the operator string is usually chosen so
long that the worm never or very rarely reaches the root of the tree.
This means that there are almost always nodes close to the root with
operators that are never updated and thus act on the trial-state after
every update.  In this way, we are effectively using an optimized
trial-state.  The effect is similar to generating the trial-state by
performing several updates on a randomly chosen trial-state and taking
the resulting state for the actual calculation.  We used such a
trial-state for the conventional VBQMC-calculations shown in this
section. 

\begin{figure}[htb]
  \includegraphics[]{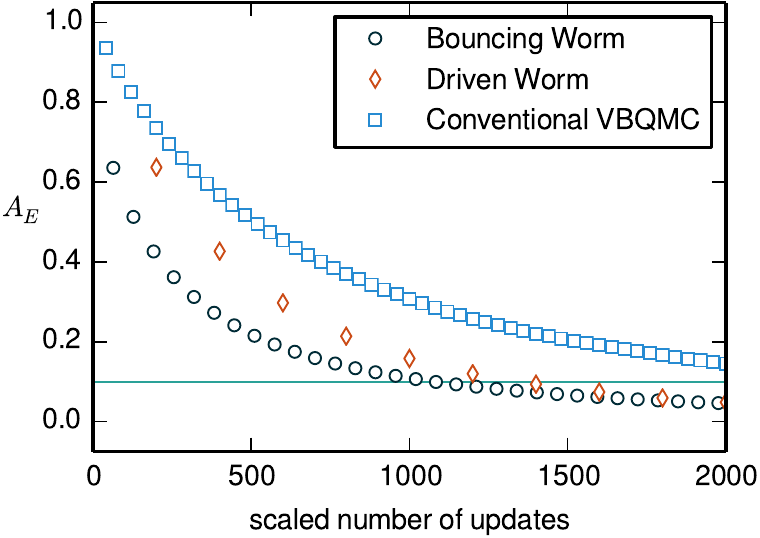}
  \caption{The autocorrelation-function versus scaled number of updates
    for the two worm algorithms and conventional VBQMC\@.  The number of
    updates was scaled by the number of operators attempted to be
    changed in an update, $\langle l/2 \rangle$.  Hence, data for
    conventional VBQMC updates, the driven worm algorithm and the
    bouncing worm algorithm were multiplied by $4$, $200$ and $63.758$,
    respectively. For the worm algorithms the same parameters as in
    Fig.~\ref{fig_comp_sivabo_driven_bounce} were used. This means that
    $\alpha=0.995$ and $b_u=0.2675$.  An operator-string of
    length 1000 was used for all three algorithms.  The horizontal line
    at $0.1$ was added to allow for easy visual estimation of the scaled
    autocorrelation-time.
  }
  \label{fig_comp_sivabo_driven_bounce_autocorfu}
\end{figure}
A useful measure of the effectiveness of an algorithm can be obtained
from the autocorrelation function. If simply measured as a function of
the number of updates it decreases dramatically faster for the worm
algorithms when compared to conventional VBQMC\@.  However, just using
one update as the temporal unit puts conventional VBQMC at an unfair
disadvantage. The reason is, that in calculations with conventional
VBQMC   one attempts to change 4 operators per update while for the worm
algorithms it could be many more. The number of updated operators in a
given worm update varies greatly with the length of the worm, $\langle
l\rangle$, which can easily be hundreds of operators long.  Since a
single worm update is, therefore, computationally more expensive to
perform than a single 4 operator update with conventional VBQMC, it
seems fairer to compare autocorrelation functions with this difference
taken into account. That is, a fair comparison would ask which algorithm
has the smallest correlations when on average the same number of changes
has been attempted. We can take this into account by simply scaling the
temporal axis with the average size of the attempted update.

In Fig.~\ref{fig_comp_sivabo_driven_bounce_autocorfu} we therefore show
results for the energy autocorrelation function for the two worm
algorithms as well as for conventional VBQMC with the temporal axis
rescaled by the number of operators one attempts to change in a single
update. During one update with worm-algorithms one tries to update
$l/2$ operators.  The scaled number of updates is 
simply $\#\mathrm{updates}\times\langle l/2 \rangle$ with $\langle l/2
\rangle=4$ for conventional VBQMC and $\langle l/2 \rangle= \langle r
\rangle = 1/(1-\alpha)$
for the driven worm algorithm. For the bouncing worm algorithm, $\langle
l \rangle$ has to be measured during the simulation, since the bouncing
worm can go up and down the tree many times. Thus, $\langle l/2 \rangle$
can be orders of magnitudes bigger than the mean penetration-depth.  For
instance, for the calculations shown in
Fig.~\ref{fig_comp_sivabo_driven_bounce_autocorfu} the mean
penetration-depth was approximately 7.8 whereas $\langle l/2 \rangle=
63.758$.  Even including such a rescaling of the temporal axis, it is
clear that the autocorrelation-times are much shorter for the worm
algorithms, as shown in
Fig.~\ref{fig_comp_sivabo_driven_bounce_autocorfu}.  

\begin{figure}[thb]
  \includegraphics[]{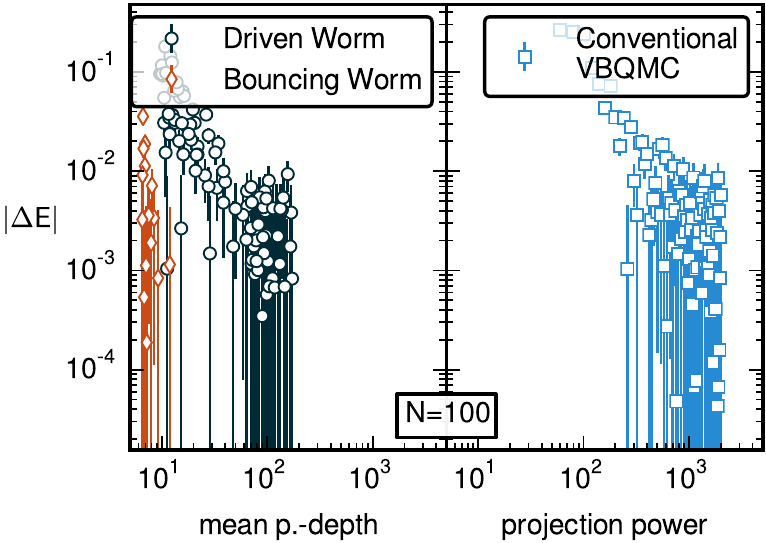}
  \caption{The deviation, $|\Delta E|$, from the exact Bethe-ansatz results
    for a chain with 100 sites. The results are shown for the driven and
    bouncing worm algorithms versus the mean penetration-depth and for
    conventional VBQMC updates versus the projective power (length of operator
    string).  The worm algorithms reach the same small value of $|\Delta
    E|$ with a mean penetration-depth an order of magnitude smaller than
    the projective power used for the calculation with conventional
    VBQMC updates.  The bars on the markers indicated the statistical
    uncertainty. The colored (dark) surfaces are due to overlapping
    error-bars.  Operator-strings of 20,000 operators were used for the
    calculations with the worm algorithms.
  }
  \label{fig_sivabo_driven_bounce_a_u}
\end{figure}
The two worm algorithms change operators of the string starting from one
end while the conventional VBQMC selects 4 operators at random to be
changed.  As mentioned in subsection~\ref{subsec_bouncing_worm}, the
mean and the maximum penetration-depth are usually much smaller than the
length of the operator-string (the projection power). It is therefore
natural to ask if one can reach a similar quality of results using worm
algorithms and conventional VBQMC\@.

That this is so can be seen by plotting the absolute deviation from the
ground-state energy, $|\Delta E|$, versus the mean penetration-depth.
As shown in Fig.~\ref{fig_sivabo_driven_bounce_a_u}, the mean
penetration-depth can, in fact, be much smaller than the projection
power of a conventional VBQMC-calculation and still yield results of the
same accuracy.  

\begin{figure}[thb]
  \includegraphics[]{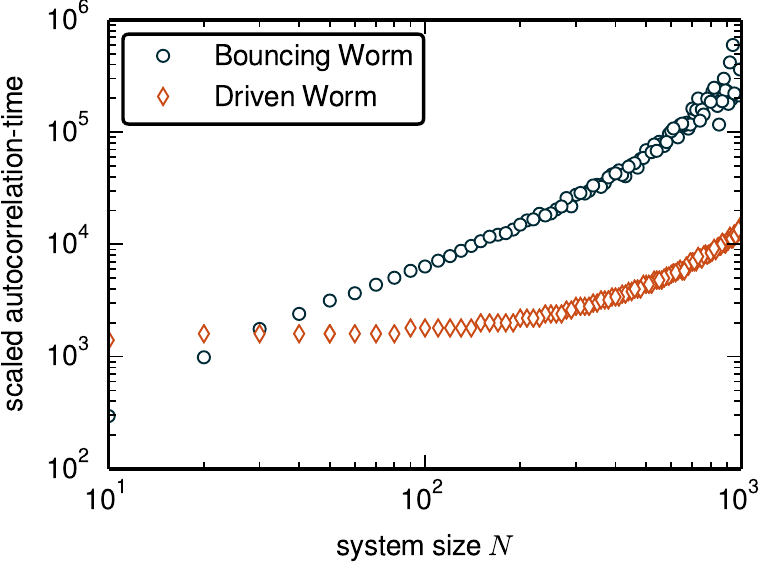}
  \caption{The scaled autocorrelation-time for the driven and bouncing
    worm algorithms as a function of system size.  A fixed operator
    string of length $20,000$ was used in the calculations.  In the
    calculations shown, the bouncing worm algorithm was run with
    $b_u=0.275$ and the driven worm algorithm was run with
    $\alpha=0.995$.  For the scaling we use $\langle l/2 \rangle =
    200$ for the driven worm algorithm and an  $\langle l/2 \rangle$
    between $60$ for $N=10$ and $221$ for $N=1000$ for the
    bouncing worm algorithm.
  }
  \label{fig_algoD_algoC_SiVaBo}
\end{figure}

Finally, we look at how the scaled autocorrelation-time depends on the
size of the system studied.  For convenience, we define the scaled
autocorrelation-time to be the point where the autocorrelation function
has decreased to the value $0.1$ (see
Fig.~\ref{fig_comp_sivabo_driven_bounce_autocorfu}). Since in realistic
calculations one would use a fixed (large) length of operator string
with the worm algorithms, while one would scale it with the size of the
system in conventional VBQMC, we here only compare the two worm
algorithms.  Our results are shown in Fig.~\ref{fig_algoD_algoC_SiVaBo}
for a fixed length operator string of $20,000$.

For the simulations shown in Fig.~\ref{fig_algoD_algoC_SiVaBo} the mean
penetration-depths for the driven worm algorithm were about $100$. The
autocorrelation-time for the driven worm algorithm starts to increase
appreciably at this system size, while it is  initially are almost flat.
We conclude that a significant increase in the autocorrelation-time
appears once the system size significantly exceeds the mean
penetration-depth. A similar effect can be observed for the bouncing
worm algorithm.  The mean penetration-depths for the bouncing worm
algorithm are, however,   much smaller (around $9$; see
Fig.~\ref{fig_sivabo_driven_bounce_a_u}). Results for $N$ smaller than
the mean penetration-depth are therefore not shown in
Fig.~\ref{fig_algoD_algoC_SiVaBo}.  The autocorrelation-times remain
manageable for the system sizes studied, even though it is consistently
increasing.

Compared to simple implementations of VBQMC, the worm algorithms have
significant overhead.  This is largely compensated by the large number
of operators that can be changed in an update and resulting shorter
autocorrelation-times, as we found in all computations.  Given the
somewhat different properties of the two worm algorithms, a realistic
implementation could combine the two by performing updates with the
driven worm algorithm mixed with updates using the bouncing worm
algorithm (and perhaps conventional VBQMC updates).

\section{Conclusion}
\label{sec_conclusion}

We have shown that valence-bond quantum Monte Carlo can be implemented
with an update build around the notion of a worm propagating through a
tree.  Many different such algorithms are possible. We studied the
validity and efficiency of two of them. One for which no update is
rejected (the bouncing worm algorithm) and one for which big parts of
the operator-string are updated (the driven worm algorithm). Both
algorithms are attractively simple and straight forward to implement
and produce high quality results. 

While they may not be computationally competitive with state of the art
loop update algorithms~\cite{sandvik_loop_2010} for VBQMC, the
algorithms presented here are intrinsically interesting since they
represent a new class of algorithms that should be generally applicable
to projective methods. These algorithms are not restricted to the
valence bond basis and preliminary results show that they can be quite
efficient in the $S^z$-basis~\cite{SzProj_2014} method and might spark
further development of it.  We also note that many other algorithms can
easily be found with the results contained in this paper and that it is
possible that the parameter space allows for much more efficient
algorithms than the two we have studied here.

In terms of further optimizing the algorithms several directions may be
interesting to pursue.  Not updating some of the operators the worm
visits, might boost the acceptance ratio of the driven worm algorithms
and thereby reduce the autocorrelation-times. This could be combined
with attempting to reduce the overhead of the driven worm calculations
by always forcing the worm all the way down to the root.  This would
eliminate the need to keep track of the state at each node.  With the
current practice of updating all operators after turning around, going
all the way to the node during every update leads to very small
acceptance ratios.

We acknowledge computing time at the Shared Hierarchical Academic
Research Computing Network (SHARCNET:www.sharcnet.ca) and research
support from NSERC.

\appendix
\section{Pseudocode for implementations of the tree worm algorithm}
This appendix contains pseudocode that shall serve to clarify the
algorithms proposed in this paper. To simplify notation we refer to
diagonal operator as DOP and non-diagonal operators as NDOP.
\subsection{Bouncing worm algorithm}
\label{subsec_bouncing_worm_pse}
In this section we give detailed information on a straightforward
(albeit not optimized) implementation of the bounce-algorithm (see
Subsec.~\ref{subsec_bouncing_worm}).  Shown is an outline of the central
part of the algorithm: the update of the operator-string and the state. 

The algorithm works its way up the tree.  It starts at the last branch
which is assigned the $n$th position. At each position it is decided if
the worm goes up the tree or down, in which case a new operator is
chosen for the branch at this position. The necessary probabilities are
calculated according to the expressions given in
Eq.~\ref{eq_algoD_weights} and Eq.~\ref{eq_bounce_c}. If a new
operator is chosen for the $n$th branch, the update is complete.

It is assumed that the tree is so high (the operator-string so long)
that the root is never reached. If the root is reached, one has to
choose an operator for the first branch according to the probabilities
outlined in Sec.~\ref{sec_tree_algorithm} after Eq.~\ref{eq_cond}. 

Schematically, using pseudocode, a bouncing worm update of a tree with
$n$ nodes can be outlined as follows:  

\begin{mdframed}[style=algstyle]
  \begin{algorithmic}
    \State $\mathtt{pos} = n$ \Comment{start at last branch}
    \State $\mathtt{going\_up} = $ TRUE
    \While {$\mathtt{pos}\;!= n+1$}
    \State  $\mathtt{ran} = \mathrm{uniform(0,1)}$
    \If {$\mathtt{going\_up}$}
    \If {operator at $\mathtt{pos}$ is NDOP} 
    \State $\mathtt{pos} = \mathtt{pos} - 1$
    \ElsIf {$\mathtt{ran} < cw$} 
    \State $\mathtt{pos} = \mathtt{pos} - 1$
    \Else 
    \State $\mathtt{going\_up} = $ FALSE 
    \State choose new DOP at $\mathtt{pos}$ 
    \State update state, $w$ and $N_1$ at $\mathtt{pos}$
    \State $\mathtt{pos} = \mathtt{pos} + 1$
  \EndIf
  \Else 
  \If{$\mathrm{ran} < b_u$}
  \State $\mathtt{going\_up} =$ TRUE

  \Else
  \If {$\mathtt{ran} - b_u < w N_1 $}
  \State choose DOP at $\mathtt{pos}$
  \Else 
  \State choose NDOP at $\mathtt{pos}$
\EndIf
\State update state, $w$ and $N_1$ at $\mathtt{pos}$
\State $\mathtt{pos} = \mathtt{pos} + 1$
  \EndIf
\EndIf
\EndWhile
\end{algorithmic}
\end{mdframed}
The weights $w, u$ and $b_u$, $N_1$ as well as the state are stored at
each node. 

\subsection{Driven worm algorithm}
\label{subsec_driven_worm_pse}
We now turn to a description of a (not optimized) implementation
of the driven worm algorithm 
(see Subsec.~\ref{subsec_driven_worm}). As above, we show an outline of the
central part of the algorithm: the update of the operator-string and the
state. 

The worm works its way up the tree. It starts at the last branch which
is assigned the position $n$.  While going up the tree, the worm, at
each node,  goes further up the tree with probability $\alpha$ or turns
around with probability $1$\hspace{0pt}$-$\hspace{0pt}$\alpha$. After
turning around, the worm keeps going down until it reaches the end. At
the nodes the worm visits new operators are chosen.  When the worm
reaches the end, it has to be decided whether or not the update should
be accepted.  The associated probabilities are calculated according to
the expressions given in the main text (see
Eq.~\ref{eq_alphadefinition}, Eq.~\ref{eq_algoC_weights} and
Eq.~\ref{eq_accprob}).

As above, we assume that the tree is so high (the operator-strings so long)
that the root is never reached.  If that the root is reached, one has to choose an
operator for the first branch according to the probabilities outlined in
Sec.~\ref{sec_tree_algorithm} after Eq.~\ref{eq_cond}. 

Shown is the driven worm update of a tree with $n$ nodes. 
Using pseudocode language, a driven worm update then takes the following form for a tree
with $n$ nodes:

\begin{mdframed}[style=algstyle]
  \begin{algorithmic}
    \State $\mathtt{pos} = n$ \Comment {start at last branch}
    \State $\mathtt{going\_up} = $ TRUE
    \While {$\mathtt{pos}\;!= n+1$}
    \State  $\mathtt{ran} = \mathrm{uniform(0,1)}$
    \While {$\mathtt{going\_up}$}
    \If {$\mathtt{ran} < \alpha $} 
    \State $\mathtt{pos} = \mathtt{pos} - 1$
    \Else 
    \State $\mathtt{going\_up} = $ FALSE 
    \If {operator at $\mathtt{pos}$ is DOP} 
    \If {$\mathtt{ran} - \alpha < y N_{1 / 2} $}
    \State choose new NDOP at $\mathtt{pos}$
    \Else
    \State choose DOP at $\mathtt{pos}$ 
  \EndIf
  \Else 
  \If {$\mathtt{ran} - \alpha < z N_{1} $}
  \State choose new DOP at $\mathtt{pos}$
  \Else
  \State choose NDOP at $\mathtt{pos}$ 
\EndIf
\State update state, weights, $c$, and $N_1$ at $\mathtt{pos}$
\State $\mathtt{pos} = \mathtt{pos} + 1$
    \EndIf
  \EndIf
\EndWhile
\If {$\mathtt{ran} < w N_1 $}
\State choose DOP at $\mathtt{pos}$
\Else 
\State choose NDOP at $\mathtt{pos}$
    \EndIf
    \State update state, weights, $c$, and $N_1$ at $\mathtt{pos}$
    \State $\mathtt{pos} = \mathtt{pos} + 1$
  \EndWhile
  \State Accept or reject using old and new $c$'s.
\end{algorithmic}
\end{mdframed}
The weights, $c$, $N_1$ as well as the state are stored at each node. A
new string is not always accepted.

\end{document}